# Erasure versus teleportation scheme of optical CNOT gate

Antoni Wójcik* and Andrzej Grudka**

Faculty of Physics, Adam Mickiewicz University,

Umultowska 85, 61-614 Poznań, Poland

Abstract

We clarify the connections between the erasure scheme of probabilistic CNOT gate implementation recently proposed by Pittman, Jacobs and Franson [Phys. Rev. A 64, 062311 (2001)] and quantum teleportation.

## I. Introduction

Many concepts of quantum information have been experimentally verified with the use of linear optics (e.g. quantum teleportation [1], quantum dense coding [2], quantum cryptography [3], decoherence-free subspaces [4], entanglement concentration [5], Schumacher compression [6]). It has been possible because qubits encoded in photonic degrees of freedom are relatively immune against decoherence and are easy to manipulate as far as single qubit operations are concerned. However, the most challenging problem is the optical realization of a scalable quantum computer which requires nonlinear interaction between two photons. Recently, it has been shown by Knill, Laflamme and Milburn [7] that this nonlinearity can be generated by suitably chosen measurements and postselection. With the use of such artificial nonlinearity two-qubit gates can be implemented in a probabilistic manner. For example CNOT gate was realized experimentally by O'Brien et al. [8] (with the probability of success equal to $1/9$) and by Pittman et al [9] (with the probability of success equal to $1/4$). It was also shown that the probability of success can be made arbitrary close to one [7, 10], however, it would require complicated quantum networks together with many auxiliary qubits. In this paper we concentrate on some simple nondeterministic schemes of quantum CNOT gate which require only a few auxiliary photons. The very first scheme of such a linear optics implementation of CNOT gate [7, 11] is based on quantum teleportation [12]. It requires four auxiliary photons in an entangled state and succeeds with a probability of $1/4$. Pittman, Jacobs and Franson (PJF) proposed an alternative scheme which gives the same probability of success requiring, however, only a single pair of entangled auxiliary photons [13]. Their scheme combines two operations - quantum parity check and quantum encoder and is based on quantum-erasure method. It was claimed by PJF that their scheme *does not rely on teleportation in any obvious way*. In this paper we would like to clarify the relations between the PJF scheme and the teleportation scheme. We will show that although the PJF scheme does not rely on teleportation in an obvious way, there exists a description in which PJF can be regarded as a quantum teleportation scheme. It should be mentioned that the elements of PJF CNOT gate, i.e. quantum parity check and destructive CNOT, have been already realized experimentally [14-16]. The paper is organized as follows: in Section II we remind the PJF erasure scheme, in Section III we remind the teleportation scheme and finally in Section IV we describe the connections between both schemes.

II. Erasure CNOT scheme

The CNOT gate implemented by erasure scheme of PJF (E-CNOT) acts on qubits encoded in the polarization states $|H\rangle$ and $|V\rangle$. It should perform the following operations

$$|H\rangle|H\rangle \to |H\rangle|H\rangle$$
$$|H\rangle|V\rangle \to |H\rangle|V\rangle$$
$$|V\rangle|H\rangle \to |V\rangle|V\rangle \qquad (1)$$
$$|V\rangle|V\rangle \to |V\rangle|H\rangle$$

The optical scheme of the E-CNOT gate is presented in Fig. 1. The gate performs nonunitary operation on polarization states of four modes, two input modes (IN, IN') and two auxiliary modes (A, A'). It consists of two parts which can be called (after PJF) quantum parity check and destructive CNOT (D-CNOT). Each of these parts can be treated as a programmable quantum gate operating on input mode, whereas the auxiliary mode controls the gate action. The key element of the scheme is the gate F. Its optical implementation is presented in Fig. 2. It consists of two polarizing beam splitters (PBS), one half-wave plate (HWP), two single-photon detectors (0 and 1). Moreover, there is an element PC (e.g. Pockels cell) performing conditional phase flip. PBS transmits the horizontally polarized photons and reflects the vertically polarized ones. The major axis of HWP is oriented at the angle 22.5º with respect to the horizontal polarization which means that it performs the following unitary operation

$$|H\rangle \to \frac{-i}{\sqrt{2}}(|H\rangle + |V\rangle)$$
$$|V\rangle \to \frac{-i}{\sqrt{2}}(|H\rangle - |V\rangle) \qquad (2)$$

Thus, the HWP performs (up to global phase) the Hadamard gate. The F gate operates in a nondeterministic way, namely we accept the operation performed on input mode only in the case when exactly one photon is detected. In the case of detection of one photon in detector A and zero photons in detector B the element PC is inactive. On the other hand, in the case of detection of one photon in detector B and zero photons in detector A, we flip the phase with the use of PC. In order to understand the action of the F gate let us assume that photon in the auxiliary mode is horizontally or vertically polarized. Of course, the only accepted cases by post-selection are those where both photons are reflected or both photons are transmitted by the first PBS. It follows that the F gate performs the operation

$$(\alpha|H\rangle + \beta|V\rangle)_{IN}|H\rangle_A \xrightarrow{F} \alpha|H\rangle_{IN}$$
$$(\alpha|H\rangle + \beta|V\rangle)_{IN}|V\rangle_A \xrightarrow{F} \beta|V\rangle_{IN} \quad . \quad (3)$$

Thus one can see why this operation is called "quantum parity check". On the other hand, if the auxiliary mode is prepared in the following superposition $|+\rangle = \frac{1}{\sqrt{2}}(|H\rangle + |V\rangle)$, then the F gate performs the "neutral filter" operation

$$(\alpha|H\rangle + \beta|V\rangle)_{IN}|+\rangle_A \xrightarrow{F} \frac{1}{\sqrt{2}}(\alpha|H\rangle + \beta|V\rangle)_{IN}, \quad (4)$$

where $1/\sqrt{2}$ indicates that the operation succeeds with the probability of $1/2$. If instead of the state $|+\rangle$ we use the state $|-\rangle = \frac{1}{\sqrt{2}}(|H\rangle - |V\rangle)$, the F gate flips the relative phase between states $|H\rangle$ and $|V\rangle$

$$(\alpha|H\rangle + \beta|V\rangle)_{IN}|-\rangle_A \xrightarrow{F} \frac{1}{\sqrt{2}}(\alpha|H\rangle - \beta|V\rangle)_{IN}. \quad (5)$$

We will now proceed to the construction of a D-CNOT gate. As is well known, the CNOT gate is equivalent to the controlled phase flip gate (CZ) sandwiched between two Hadamard gates. In the optical scheme considered (see Fig. 1) the Hadamard gates are implemented by HWP$_{1,2}$. Thus, we only have to show that F gate together with the Hadamard gate acting on the auxiliary mode (HWP$_3$) performs the controlled phase flip in the input mode. Because the state $|H\rangle$ ($|V\rangle$) of the auxiliary mode is transformed by the HWP$_3$ into $|+\rangle$ ($|-\rangle$), Eqs. (4, 5) imply that HWP$_3$ together with F perform the CZ gate. Thus, the action of D-CNOT provided that the auxiliary mode is in the state $|H\rangle$ or $|V\rangle$ is given by the formulae

$$(\gamma|H\rangle + \delta|V\rangle)_{IN'}|H\rangle_{A'} \xrightarrow{D-CNOT} \frac{1}{\sqrt{2}}(\gamma|H\rangle + \delta|V\rangle)_{IN'}$$
$$(\gamma|H\rangle + \delta|V\rangle)_{IN'}|V\rangle_{A'} \xrightarrow{D-CNOT} \frac{1}{\sqrt{2}}(\gamma|V\rangle + \delta|H\rangle)_{IN'} \quad (6)$$

The D-CNOT gate presented in Fig. 1 is destructive in the sense that the controlled qubit (polarization state of the auxiliary mode) is destroyed. Let us now show how PJF combine the quantum parity check with the D-CNOT in order to perform full CNOT gate on two input modes. If both auxiliary modes are prepared in the same polarization state then obviously one gets

$$(\alpha|H\rangle + \beta|V\rangle)_{IN}(\gamma|H\rangle + \delta|V\rangle)_{IN'}|H\rangle_A|H\rangle_{A'} \xrightarrow{E-CNOT} \frac{1}{\sqrt{2}}\alpha|H\rangle_{IN}(\gamma|H\rangle + \delta|V\rangle)_{IN'}$$
$$(\alpha|H\rangle + \beta|V\rangle)_{IN}(\gamma|H\rangle + \delta|V\rangle)_{IN'}|V\rangle_A|V\rangle_{A'} \xrightarrow{E-CNOT} \frac{1}{\sqrt{2}}\beta|V\rangle_{IN}(\gamma|V\rangle + \delta|H\rangle)_{IN'} \quad (7)$$

Thus if auxiliary modes are prepared in a maximally entangled state $|\Phi\rangle_{AA'} = \frac{1}{\sqrt{2}}(|H\rangle_A|H\rangle_{A'} + |V\rangle_A|V\rangle_{A'})$ then the CNOT gate is performed on the input modes i.e.

$$(\alpha|H\rangle + \beta|V\rangle)_{IN}(\gamma|H\rangle + \delta|V\rangle)_{IN'}|\Phi\rangle_{AA'} \xrightarrow{E-CNOT}$$
$$\to \frac{1}{2}(\alpha\gamma|H\rangle_{IN}|H\rangle_{IN'} + \alpha\delta|H\rangle_{IN}|V\rangle_{IN'} + \beta\gamma|V\rangle_{IN}|V\rangle_{IN'} + \beta\delta|V\rangle_{IN}|H\rangle_{IN'}) \quad (8)$$

The one half factor indicates that the operation succeeds with the probability of $1/4$. Let us emphasize that the scheme presented in Fig. 1 is equivalent to the scheme proposed by PJF (see Fig. 6 in [13]). The only difference (introduced here for pedagogical reason) is that instead of rotating the polarizing beam splitters we rotate polarization of photons with the use of half-wave plates.

## II. Teleportation CNOT scheme

As shown by Gottesman and Chuang [12], quantum gates can be implemented with the use of quantum teleportation. The scheme of such a gate (telegate T) is presented in Fig. 3.

The telegate operates on the input mode. The state of auxiliary qubits encodes the operation which is to be performed. The basic element of the telegate is the measurement in the Bell basis. We will later assume that our qubits will be encoded in some photonic degrees of freedom. Thus, only partial Bell measurement (PBM), which enables a distinction between two of four Bell states, namely $|\Psi^{\pm}\rangle = (|0\rangle|1\rangle \pm |1\rangle|0\rangle)/\sqrt{2}$, will be considered. Secondly, the correction $Z^j$ is applied according to the result $j$ of PBM, where $j = 0\,(1)$ stands for $|\Psi^+\rangle$ ($|\Psi^-\rangle$). Z is the operator of the phase flip ($Z = \sum_{k=0}^{1} (-1)^k |k\rangle\langle k|$). Let us consider the action of the telegate in two simple cases, namely when the auxiliary modes are in the $|\Psi^+\rangle$ or $|\Psi^-\rangle$ states. The state of the whole system (three modes) can be written as

$$|\varphi\rangle_{IN} |\Psi^{\pm}\rangle_{A1A2} = $$
$$= \frac{1}{2}\left( |\Psi^{\pm}\rangle_{INA2} |\varphi\rangle_{A1} + |\Psi^{\mp}\rangle_{INA2} (Z|\varphi\rangle_{A1}) + |\Phi^{\pm}\rangle_{INA2} (X|\varphi\rangle_{A1}) + |\Phi^{\mp}\rangle_{INA2} (ZX|\varphi\rangle_{A1}) \right), \quad (9)$$

where $|\varphi\rangle_{IN}$ is an arbitrary state of the input qubit, Bell states $|\Phi^{\pm}\rangle = (|0\rangle|1\rangle \pm |1\rangle|0\rangle)/\sqrt{2}$ and $X = \sum_{k=0}^{1} |1-k\rangle\langle k|$. By swapping the modes IN and A1, the state is transformed into

$$\frac{1}{2}\left( |\varphi\rangle_{IN} |\Psi^{\pm}\rangle_{A1A2} + (Z|\varphi\rangle_{IN}) |\Psi^{\mp}\rangle_{A1A2} + (X|\varphi\rangle_{IN}) |\Phi^{\pm}\rangle_{A1A2} + (ZX|\varphi\rangle_{IN}) |\Phi^{\mp}\rangle_{A1A2} \right). \quad (10)$$

Now, the partial Bell measurement is performed on auxiliary modes A1 and A2 and correction $Z^j$ is applied. Only $|\Psi^{\pm}\rangle$ are accepted results of this measurement. Thus the action of the telegate T can be written as

$$|\varphi\rangle_{IN} |\Psi^+\rangle_{A1A2} \xrightarrow{T} \frac{1}{\sqrt{2}} |\varphi\rangle_{IN}$$
$$|\varphi\rangle_{IN} |\Psi^-\rangle_{A1A2} \xrightarrow{T} \frac{1}{\sqrt{2}} Z|\varphi\rangle_{IN} \quad . \quad (11)$$

It seems to look little funny to propose a special method for performing identity or simple Z gate with probability $1/2$. However, as shown in [12] two telegates combined allow one to perform a controlled Z gate ($CZ = \sum_{m,k=0}^{1}(-1)^{km}|m\rangle|k\rangle\langle k|\langle m|$). To see this let us consider the action of two telegates for different combinations of $|\Psi^{\pm}\rangle$ as auxiliary qubits. From Eqs. (11) we have

$$
\begin{aligned}
|\varphi\rangle_{IN}|\chi\rangle_{IN'}|\Psi^{+}\rangle_{A1A2}|\Psi^{+}\rangle_{A1'A2'} &\xrightarrow{T\otimes T} \frac{1}{2}|\varphi\rangle_{IN}|\chi\rangle_{IN'} \\
|\varphi\rangle_{IN}|\chi\rangle_{IN'}|\Psi^{+}\rangle_{A1A2}|\Psi^{-}\rangle_{A1'A2'} &\xrightarrow{T\otimes T} \frac{1}{2}|\varphi\rangle_{IN}(Z|\chi\rangle)_{IN'} \\
|\varphi\rangle_{IN}|\chi\rangle_{IN'}|\Psi^{-}\rangle_{A1A2}|\Psi^{+}\rangle_{A1'A2'} &\xrightarrow{T\otimes T} \frac{1}{2}(Z|\varphi\rangle)_{IN}|\chi\rangle_{IN'} \\
|\varphi\rangle_{IN}|\chi\rangle_{IN'}|\Psi^{-}\rangle_{A1A2}|\Psi^{-}\rangle_{A1'A2'} &\xrightarrow{T\otimes T} \frac{1}{2}(Z|\varphi\rangle)_{IN}(Z|\chi\rangle)_{IN'}
\end{aligned} \quad (12)
$$

CZ can be decomposed in the following way

$$CZ = I\otimes I + I\otimes Z + Z\otimes I - Z\otimes Z. \quad (13)$$

All terms of the above equation can be realized (with probability $1/4$) by teleportation as is seen from Eqs. (12). As follows from the linearity of $T\otimes T$, one has to take appropriate superposition of the auxiliary qubits in order to perform CZ gate. Eqs. (12) and (13) give the following form of this superposition

$$
\begin{aligned}
|\Psi^{CZ}\rangle_{A1A2A1'A2'} = \\
= |\Psi^{+}\rangle_{A1A2}|\Psi^{+}\rangle_{A1'A2'} + |\Psi^{+}\rangle_{A1A2}|\Psi^{-}\rangle_{A1'A2'} + |\Psi^{-}\rangle_{A1A2}|\Psi^{+}\rangle_{A1'A2'} - |\Psi^{-}\rangle_{A1A2}|\Psi^{-}\rangle_{A1'A2'}
\end{aligned}, \quad (14)
$$

or equivalently

$$
\begin{aligned}
|\Psi^{CZ}\rangle_{A1A2A1'A2'} = \\
= |0\rangle_{A1}|1\rangle_{A2}|0\rangle_{A1'}|1\rangle_{A2'} + |0\rangle_{A1}|1\rangle_{A2}|1\rangle_{A1'}|0\rangle_{A2'} + |1\rangle_{A1}|0\rangle_{A2}|0\rangle_{A1'}|1\rangle_{A2'} - |1\rangle_{A1}|0\rangle_{A2}|1\rangle_{A1'}|0\rangle_{A2'}
\end{aligned}. \quad (15)
$$

Finally the CZ gate realized by two telegates T can be used in the circuit presented in Fig. 4 to perform the CNOT gate (CZ gate sandwiched between two Hadamard gates).

For the reason which becomes clear in the next section let us now present a slightly different realization of a telegate which is equivalent to the telegate T. This new telegate T' (see Fig. 5) is equivalent to T provided that we restrict our analysis to that with a combination of $|\Psi^\pm\rangle$ as auxiliary qubits (as e.g. in Eq.(14)). The main difference between T' and T is that in the place of SWAP gate (acting on modes IN and A1) the parity filter ($PF = |0\rangle|0\rangle\langle 0|\langle 0| + |1\rangle|1\rangle\langle 1|\langle 1|$) is used. Of course there is no more need to perform swap operation as in the subspace spanned by $|0\rangle|0\rangle$ and $|1\rangle|1\rangle$ it is just identity. Moreover, one does not have to worry about the states "absorbed" by PF because these states lead to unresolved (by the linear optics means) states $|\Phi^\pm\rangle$. The action of PF can be written as

$$|\varphi\rangle_{IN}|\Psi^\pm\rangle_{A1A2} \xrightarrow{PF} |\varphi\rangle_{IN}|\Psi^\pm\rangle_{A1A2} + (Z|\varphi\rangle)_{IN}|\Psi^\mp\rangle_{A1A2}. \qquad (16)$$

The equivalence of T and T' is clearly seen from the above formula compared with Eq.(10).

IV. The equivalence of the erasure and teleportation schemes

We will now redefine computational basis and analyze again the scheme presented in Fig. 2 from a different point of view. Until now, the polarization modes H and V have represented logical values 0 and 1, respectively. In this section we use mixed basis (MB), namely one computational basis for input photons and another one for auxiliary photons. For input photons we keep the polarization basis i.e. $|H\rangle_{IN} = |0\rangle_{IN}$ and $|V\rangle_{IN} = |1\rangle_{IN}$. For auxiliary photons, however, the logical values will be identified with the occupation number of a single mode. In this new basis the states $|H\rangle_A$ and $|V\rangle_A$ of a horizontally polarized photon in a spatial mode A are written as

$$|H\rangle_A = |0\rangle_{AV}|1\rangle_{AH}$$

$$|V\rangle_A = |1\rangle_{AV}|0\rangle_{AH}. \qquad (17)$$

The key point is that the single-photon states $|+\rangle_A$ and $|-\rangle_A$ when written in this basis become maximally entangled Bell states, namely

$$|+\rangle_A = \frac{1}{\sqrt{2}}\left(|0\rangle_{AV}|1\rangle_{AH} + |1\rangle_{AV}|0\rangle_{AH}\right) = |\Psi^+\rangle_{AV,AH}$$

$$|-\rangle_A = \frac{1}{\sqrt{2}}\left(|0\rangle_{AV}|1\rangle_{AH} - |1\rangle_{AV}|0\rangle_{AH}\right) = |\Psi^-\rangle_{AV,AH}$$

(18)

As a matter of fact, this kind of single-photon entanglement has been proposed in the context of quantum teleportation [17] and it has been recently implemented experimentally [18, 19]. Let us show how F gate acts on a photon in an arbitrary state $|\varphi\rangle = \alpha|0\rangle + \beta|1\rangle$ in the input mode, when the photon in the auxiliary mode is in the state $|+\rangle$ or $|-\rangle$. In MB the action of F gate is given by

$$|\varphi\rangle_{IN}|\Psi^+\rangle_{AH,AV} \xrightarrow{F} \frac{1}{\sqrt{2}}|\varphi\rangle_{IN}$$

$$|\varphi\rangle_{IN}|\Psi^-\rangle_{AH,AV} \xrightarrow{F} \frac{1}{\sqrt{2}}Z|\varphi\rangle_{IN}.$$

(19)

One sees that above equations are identical to Eqs. (11). Below we show that the F gate can be indeed understood as a telegate T'. The MB computational scheme of F gate is presented in Fig. 6. It should be emphasized that we will consider only highly restricted sets of input and output states. The initial conditions together with post-selection allow us to assume that modes AV and AH together contain only one photon. With this restriction the action of PBS given in MB by

$$(a|0\rangle + b|1\rangle)_{IN}(A|01\rangle + B|10\rangle)_{AVAH} \xrightarrow{PBS} (aA|001\rangle + bB|110\rangle)_{INAVAH} \quad (20)$$

is really equivalent to PF acting on modes IN and A1. The action of HWP described in MB is the following

$$|01\rangle_{AVAH} \xrightarrow{HWP} \frac{1}{\sqrt{2}}(|01\rangle + |10\rangle)_{AVAH}$$

$$|10\rangle_{AVAH} \xrightarrow{HWP} \frac{1}{\sqrt{2}}(|01\rangle - |10\rangle)_{AVAH}$$

(21)

The appropriate circuit built from standard gates is presented in Fig. 6. It can be used (together with detectors) to perform partial Bell state measurement according to

$$\begin{aligned}|\Psi^+\rangle &\xrightarrow{HWP} |01\rangle \\ |\Psi^-\rangle &\xrightarrow{HWP} |10\rangle\end{aligned}. \quad (22)$$

Thus, the equivalence between F gate and telegate T' can be introduced with the use of the appropriate computational basis. Finally, in order to look for equivalence of the complete schemes of CNOT implementation one has to compare the auxiliary states which feed the E-CNOT scheme and T-CNOT scheme. Let us now treat $HWP_3$ in the scheme presented in Fig. 1 not as a part of D-CNOT but as a part of auxiliary state preparation process. In this way the similarity between schemes from Figs 1 and 4 is revealed. The state of the auxiliary qubits in the E-CNOT scheme must be given by

$$(I \otimes HWP)|\Phi\rangle_{AA'} = \frac{1}{2}(|HH\rangle + |HV\rangle + |VH\rangle - |VV\rangle)_{AA'}. \quad (23)$$

It is easy to check that the above state when written in MB is equal to the auxiliary qubits state of the T-CNOT scheme given by Eq. (15).

In conclusion, we have shown how an erasure scheme of CNOT gate can be connected to a teleportation scheme. This connection becomes apparent when one re-describes the scheme in a specifically chosen basis. Our analysis does not, however, change the obvious fact that the scheme proposed by PJF is more effective in the sense that it needs as an auxiliary system only a single pair of entangled photons.

We would like to thank the Polish Committee for Scientific Research for financial support under grant no. 0 T00A 003 23.

*Email address: antwoj@amu.edu.pl

**Email address: agie@amu.edu.pl

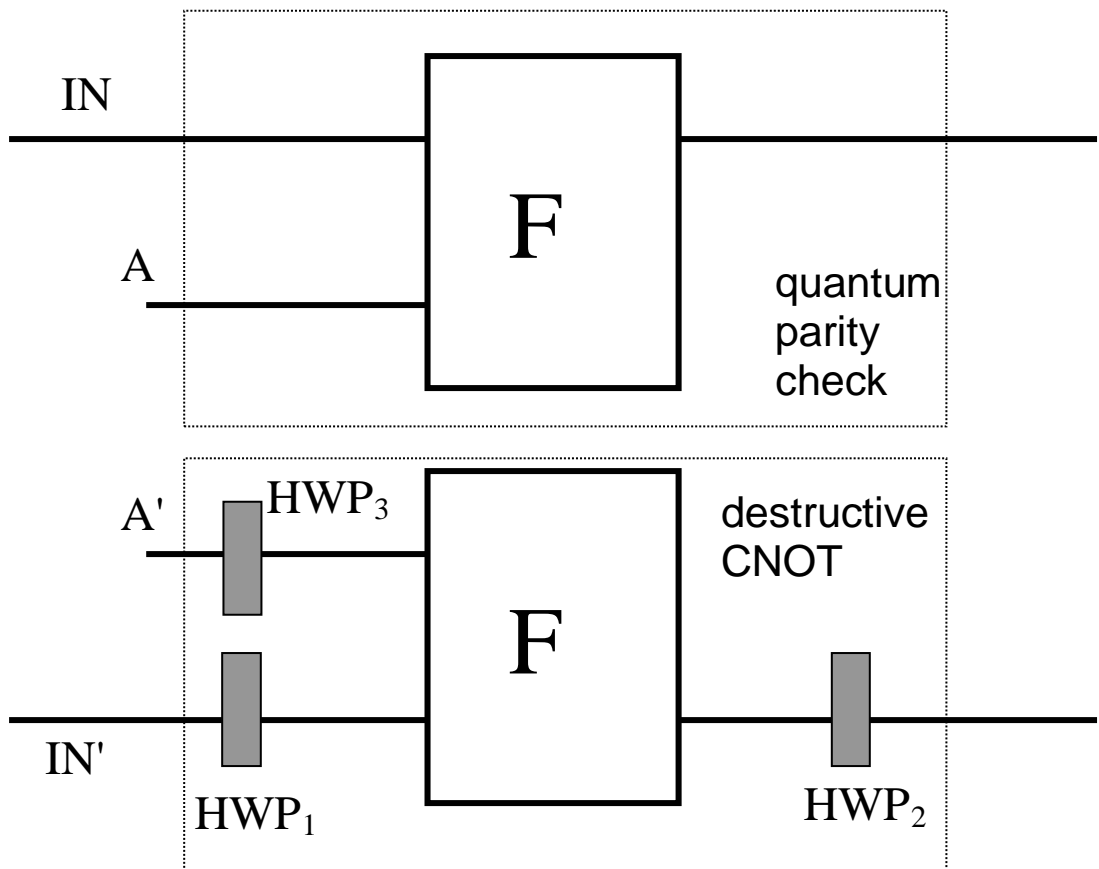

Fig. 1 Erasure scheme of CNOT gate consisting of quantum parity check and destructive CNOT (HWP - half-wave plate).

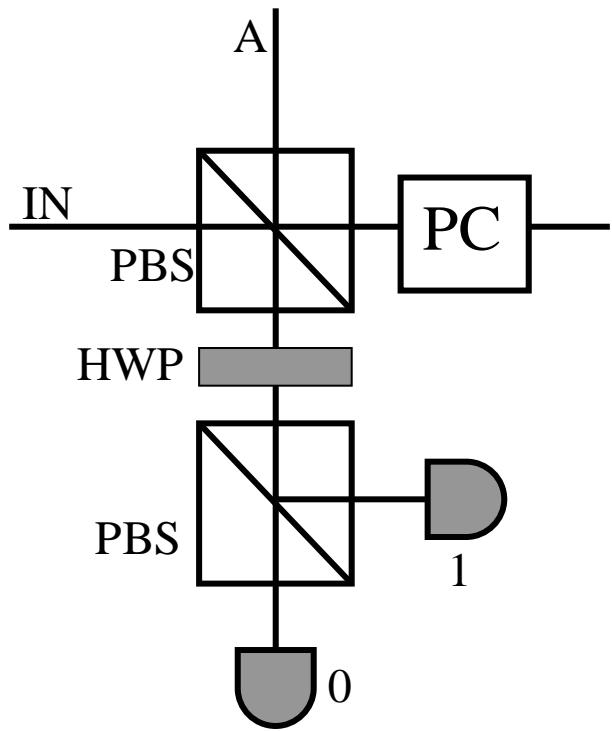

Fig. 2 Optical implementation of the F gate (PBS - polarizing beam splitter, HWP - half-wave plate, PC - Pockels cell).

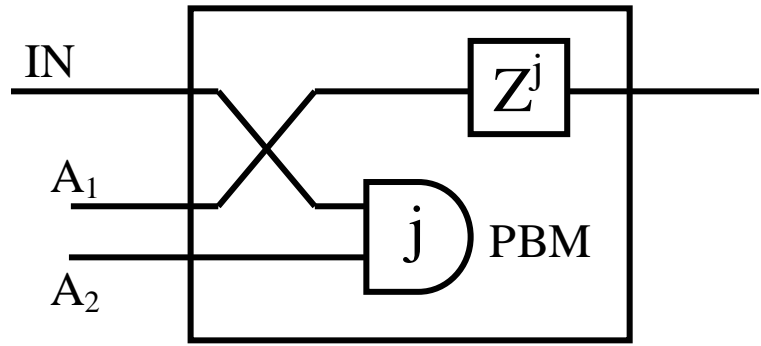

Fig. 3 Teleportation gate T (PBM - partial Bell measurement, Z - phase flip gate).

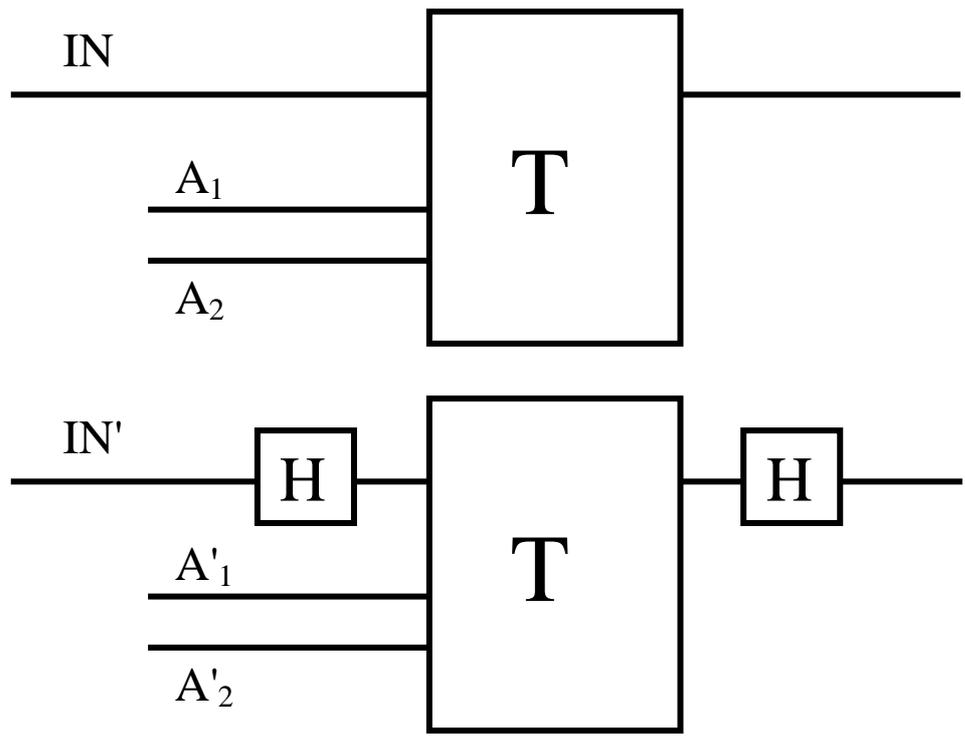

Fig. 4  Teleportation scheme of CNOT gate (H - Hadamard gate).

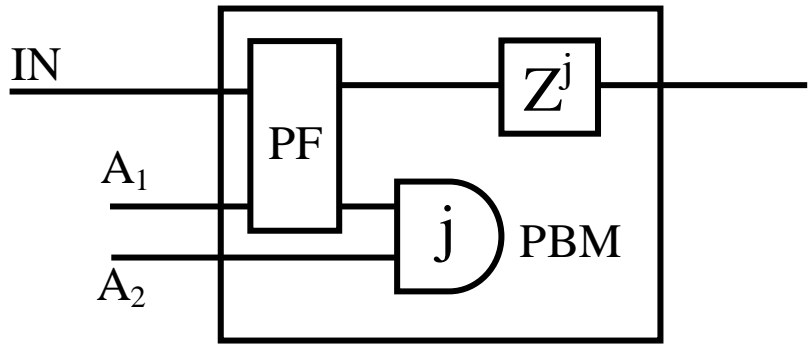

Fig. 5  Alternative teleportation scheme - gate T' (PBM - partial Bell measurement, Z - phase flip gate, PF - parity filter).

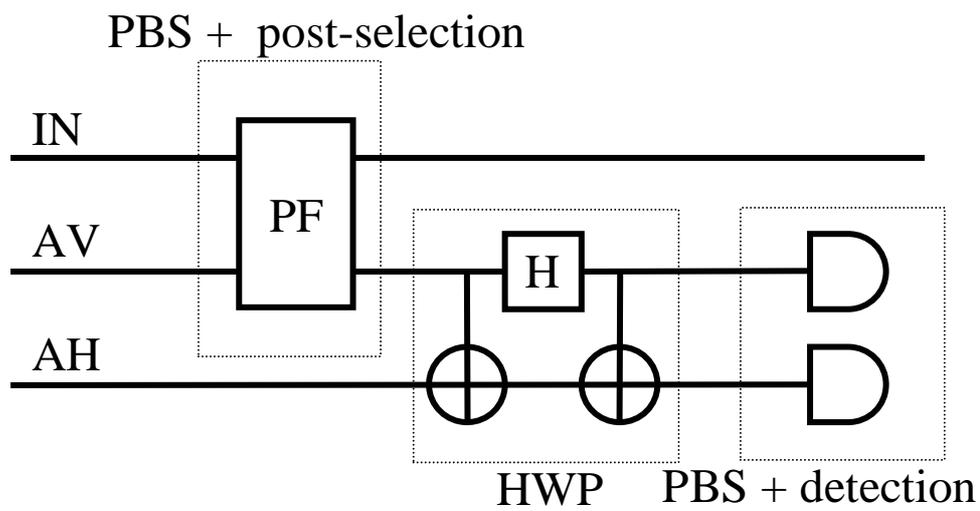

Fig. 6  Quantum circuit for gate F (PBS - polarizing beam splitter, HWP - half-wave plate, H - Hadamard gate, PF - parity filter).